\documentclass[prc,preprint,showpacs,showkeys,nofootinbib]{revtex4}%
\usepackage{amssymb}
\usepackage{graphicx}
\usepackage{bm}%
\usepackage{amsmath}%
\usepackage{amsfonts}

\newcommand{\beq}{\begin{equation}}
\newcommand{\eeq}{\end{equation}}
\newcommand{\bea}{\begin{eqnarray}}
\newcommand{\eea}{\end{eqnarray}}
\newcommand{\ben}{\begin{eqnarray*}}
\newcommand{\een}{\end{eqnarray*}}

\begin{document}
\title{Coupling of giant resonances to soft E1 and E2 modes in $^{8}$B}
\author{C.A. Bertulani}
\email{bertulani@nscl.msu.edu}
\affiliation{National Superconducting Cyclotron Laboratory, Michigan State University, East
Lansing, MI\ 48824, USA}
\date{\today }

\begin{abstract}
The dynamic coupling between giant resonance states and
\textquotedblleft soft\textquotedblright\, low-energy excitation,
modes in weakly-bound nuclei is investigated. A coupled-channels
calculation is reported for the reaction
$^{8}\mathrm{B}+\mathrm{Pb}\longrightarrow
\mathrm{p}+^{7}\mathrm{Be}+\mathrm{Pb}$ at 83 MeV/nucleon. It is
shown that the low-energy response is only marginally modified by
transitions to the isovector giant dipole and isoscalar giant
quadrupole resonances.

\end{abstract}
\maketitle

The nuclear response to external \ electromagnetic fields is one
of the main probes of the structure of nuclei far from the
stability valley\ \cite{BHM02}. The Coulomb excitation of rare
isotopes in high energy collisions ($E_{Lab}\gtrsim50$
MeV/nucleon) has revealed the existence of
\ \textquotedblleft soft\textquotedblright\ excitation modes \cite{Ie93,Zin97}%
. \ However, it is still an open question if these modes represent
a new type of collective motion, or a resonance in the continuum,
as predicted by some theories \cite{AGB}. These soft modes can
also been explained as a simple consequence of the phase-space
availability of transitions from a bound-state to a structureless
continuum \cite{BB86a}. These questions are of extreme relevance
for experimental strategies, since the Coulomb dissociation method
has become a powerful experimental alternative to access
information on the radiative capture processes occurring in
numerous astrophysical scenarios \cite{BBR86,BB88}.

It is not known if the low energy peak in the Coulomb excitation
cross sections of $^{11}\mathrm{Li}$ \cite{Zin97} is due to the
existence of a resonant state close to threshold. But, the Coulomb
breakup of $^{8}\mathrm{B}$ is well explained by a direct
transition of a bound particle state ($p_{3/2}$\ proton) to a
structureless continuum. This transition is caused by the action
of the time-dependent electric dipole (E1) and electric quadrupole
(E2) fields of the target nucleus. The resonance state at 630 keV
in $^{8}\mathrm{B}$, which plays an important role in magnetic
dipole (M1) transitions\ of\ the radiative capture reaction $^{7}%
\mathrm{Be}\left(  \mathrm{p},\gamma\right)  ^{8}\mathrm{B}$ occurring in the
sun, is imperceptible in Coulomb dissociation experiments \cite{CB96}.

The E1 and E2 response in $^{8}\mathrm{B}$ is reasonably well
described by a proton + $^{7}\mathrm{Be}$-core model with an
spectroscopic factor close to unity \cite{EB96}. This model yields
an astrophysical S-factor $S_{17}=18$ eV.b at $E_{17}=20$ keV,
where $E_{17}$ is the relative energy of the proton-beryllium
system in the solar environment. This value of $S_{17}$ is the
most recommended value, based on the average of numerous direct
and indirect experiments \cite{Da01}.

Thus, it seems that the low-energy response in $%
^{8}\mathrm{B}$\ is due entirely to the promotion of a valence
proton from the $p_{3/2}$ level into the continuum  \cite{KPK87}.
This excitation process decouples from the higher energy
excitations, except in a situation where multi-step processes are
relevant. Hence, with no configuration mixing, the electromagnetic
responses coincide with the free response (with no residual
interaction) in the low energy region. Although more elaborate
models exist in the literature \cite{PD99,Bar99}, I will adopt the
proton + $^{7}\mathrm{Be}$-core model to obtain the E1 and E2
low-energy response of $^{8}$B ($E_{x}\lesssim 5$ MeV).

Giant resonances (GRs) are collective vibrations in nuclei and
have been known for a long time (for a review, see ref.
\cite{SW91}). Their energies and widths have been studied for a
large number of nuclei. Some experimental data have been obtained
with real photons, which are well suited for (isovector) E1
excitation modes. Isoscalar E2 and higher modes have been best
studied with $\alpha$ or proton scattering. E0 (breathing) modes
have also been studied with electron and nucleus-nucleus
scattering. Nowadays a great effort is underway \cite{Au98} to
understand the structure of double giant resonances, i.e., giant
resonances excited on another giant resonance state \cite{BB86}.
Such studies have been performed with stable nuclear species. We
have practically no information about giant resonances in very
light nuclei (e.g. $^{9}\mathrm{Be}$), or in neutron or
proton-rich nuclei.(e.g., $^{8}\mathrm{B}$, or $^{11}\mathrm{Li}$)
although theoretically one expects them to exist .

The effect of continuum-continuum transitions on the low-energy
response of weakly-bound nuclei was first mentioned and studied in
ref. \cite{BC92}. More recently, intensive theoretical studies
have been performed \cite{CB96,EB96,TNT01} to access the relevance
of continuum-continuum transitions in the breakup reactions of
$^{8}\mathrm{B,}$ $^{11}\mathrm{Li}$, and other exotic light
nuclei. But besides the low energy continuum-continuum couplings,
the giant resonances located at much higher energies could also
have some influence on the low-lying states through a dynamic
coupling during the reaction process. This assumption is based on
the known fact that the giant resonances exhaust the largest part
of the electromagnetic response in heavy stable nuclei (see, e.g.,
\cite{BM75}). This often leads to a large excitation cross section
of giant resonance states in Coulomb excitation at high bombarding
energies \cite{BB88}. This hypotheses is worth investigation in
the case of light- neutron- or proton-rich nuclei.

In this article I report a study of the influence of the GR states
on the soft modes. A continuum discretized coupled-channels
calculation (CDCC) was done which includes nuclear and Coulomb
induced breakup of $^{8}\mathrm{B}$ projectiles incident on heavy
(large-Z) targets. A microscopic description of the GRs in very
light nuclei, using e.g., the random phase approximation, probably
leads to unreliable results. Thus, a more conservative approach is
adopted, describing the giant dipole resonance (GDR, $\lambda=1$)
and the giant quadrupole (isoscalar) resonance (GQR, $\lambda=2$)
by means of a Breit-Wigner function,
\begin{equation}
f_{E\lambda}(E)=\frac{C_{\lambda}}{\left(  E-E_{\lambda}\right)  ^{2}%
+\Gamma_{\lambda}^{2}/4},\label{fl}%
\end{equation}
centered on the energy $E_{\lambda}$ of the resonance. The
continuum is discretized with an energy mesh around the
resonances, using eq. \ref{fl} as reference. In terms of
$f_{E\lambda}(E)$, the total response function is given by
\begin{equation}
B\left(  E\lambda\right)  =\sum_{k}f_{E\lambda}(E_{k})\Delta E_{k}\label{BEL}%
\end{equation}
where $E\lambda=E1,$ $E2$, and $\Delta E_{k}=E_{k}-E_{k-1}$\ \ is the energy
interval. The reduced matrix elements for the excitation of the energy state
at $E=E_{k}$ from the ground state, or from a low-lying state in the continuum
($E_{i}\leq5$ MeV), are given by $\left\langle k\left\Vert O(E\lambda
)\right\Vert i\right\rangle =\left(  2I_{i}+1\right)  \sqrt{f_{E\lambda}%
(E_{k})\Delta E_{k}}$, where $I_{i}$ is the spin of the initial
state $i $. Typical values were adopted for their widths,
$\Gamma_{\lambda=1,2}=4$ MeV, and for their energy centroids,
$E_{\lambda=1}=30$ MeV and $E_{\lambda=2}=20$ MeV, respectively.

The constants $C_{\lambda}$ are obtained by assuming that the GRs exhaust
100\% of the energy-weighted sum-rule. This yields
\begin{equation}
C_{1}=\frac{9}{16\pi^{2}}\left(  2I_{i}+1\right)  \frac{\Gamma_{1}}%
{E_{\lambda=1}}\frac{\hbar^{2}}{m_{N}}\frac{NZ}{A}e^{2},\ \ \ \ \ \mathrm{and}%
\ \ \ C_{2}=\frac{15}{8\pi^{2}}\left(  2I_{i}+1\right)  \frac{\Gamma_{2}%
}{E_{\lambda=2}}\frac{\hbar^{2}}{m_{N}}R_{m}^{2}\frac{Z^{2}}{A}e^{2}%
,\label{C1C2}%
\end{equation}
where $N$, $Z,$ and $A$ are the neutron, charge, and mass numbers of the
excited nucleus, $m_{N}$ is the nucleon mass, and $R_{m}=\sqrt{\left\langle
r^{2}\right\rangle }$ is the\ ground-state density matter radius. For $^{8}B$
the value $R_{m}=2.38$ fm, obtained by a Skyrme-Hartree-Fock calculation
\cite{Br98}, is used.

The values of the normalization constants in eq. \ref{C1C2} follow from the
energy weighted-sum-rules (EWSR):%
\begin{equation}
S_{E\lambda}=\int B\left(  E\lambda;\ E\right)  dE=\int
f_{E\lambda }(E_{x})dE_{x}=\left\{
\begin{array}
[c]{c}%
\dfrac{9}{4\pi}\dfrac{\hbar^{2}}{2m_{N}}\dfrac{NZ}{A}\ {\normalsize e}%
^{2}\ \ \ \mathrm{for}\ \ \ \ E1\ \ \mathrm{isovector}\ \mathrm{excitations}\\
\dfrac{30}{4\pi}\dfrac{\hbar^{2}}{2m_{N}}\left\langle {\normalsize r}%
^{2}\right\rangle \dfrac{Z^{2}}{A}\ {\normalsize e}^{2}\ \ \mathrm{for}%
{\normalsize \ }\ \ \ E2\ \ \mathrm{isoscalar\ excitations}%
\end{array}
\right.  .\label{EWSR}%
\end{equation}
\begin{figure}
[t]
\begin{center}
\includegraphics[
height=3.179in,
width=2.4967in
]%
{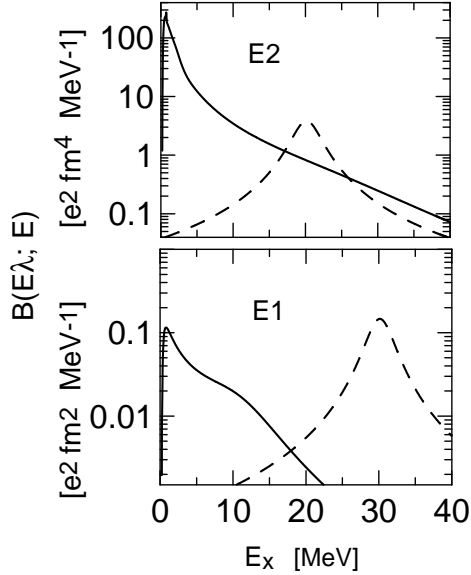}%
\caption{E2 (upper figure) and E1 (lower figure) response in $^{8}B$. The low
energy response (solid curves) was calculated with a potential model. The high
energy response (dashed curves) for the giant resonances was parametrized in
terms of Breit-Wigner functions.}%
\end{center}
\end{figure}

The contribution of nuclear excitations have been included using
the collective vibration model \cite{BM75,Satchler,Sa87}. In this
model the structure input are the deformation parameters $\delta
_{1}$ and $\delta _{2}$, obtained from the EWSR, eq. \ref{EWSR}.
That is,
\begin{equation}
\delta_{1}^{2}=\pi\frac{\hbar^{2}}{2m_{N}}\frac{A}{NZ}\frac{1}{E_{GDR}%
},\ \ \ \ \ \ \mathrm{and}\ \ \ \ \ \delta_{2}^{2}=\frac{20\pi}{3}\frac
{\hbar^{2}}{m_{N}}\frac{1}{AE_{GQR}}.\label{delta2}%
\end{equation}
To account for the width of the GRs, eq. \ref{delta2} is
multiplied by $f_{\lambda}$ given by eq. \ref{fl}, with
$C_{\lambda}=\Gamma_{\lambda}/2\pi$. In this way, the deformation
parameters acquire an energy dependence.

In the collective vibration model the nuclear excitation is induced by the
deformed potentials%
\begin{equation}
U_{1}(r,E)=\delta_{1}\left(  E\right)  \frac{3}{2}\frac{\Delta R}{R_{m}%
}\left(  \frac{dU_{opt}}{dr}+\frac{1}{3}R_{m}\frac{d^{2}U_{opt}}{dr^{2}%
}\right)  ,\ \ \ \ \ \ \mathrm{and}\ \ \ \ U_{2}(r,E)=\delta_{2}\left(
E\right)  \frac{dU_{opt}}{dr}\ ,\ \label{Uopt}%
\end{equation}
where $\Delta R=R_{p}-R_{n}=\left(  2.54-2.08\right)  $ \ fm$\ =0.46$ fm is
the difference between the proton, $R_{p}$, and the neutron, $R_{n}$, radius
in $^{8}\mathrm{B}$ \cite{Br98}. \ The optical potentials $U_{opt}$\ \ are
constructed from the ground state densities of the colliding nuclei and the
\textquotedblleft t-$\rho\rho$\textquotedblright\ approximation, as explained
in ref. \cite{BCG02}.

The electromagnetic matrix elements for the excitation of soft E1
and E2 modes were calculated with the potential model of a proton
+ $^{7}\mathrm{Be}$-core, as presented in ref. \cite{CB96}. The
2$^{+}$ ground state of $^{8}\mathrm{B}$ is described as a
p$_{3/2}$ proton coupled to a 3/2$^{-}$ ground state of the
$^{7}$Be core. The E1 soft excitations consist of transitions from
the ground state to s$_{1/2}$, d$_{3/2}$ and d$_{5/2}$ continuum
single-particle states. The E2 excitations consist of transitions
to p$_{1/2}$, p$_{3/2}$, f$_{5/2}$ and f$_{7/2}$ states.
Continuum-continuum transitions between the low-lying states have
been considered in ref. \cite{CB96} and are not taken into account
here, as we want to isolate the effect of the giant resonances.
The form of the nuclear response for the soft-modes within this
model, and for the GRs according to the parametrization described
by eqs. \ref{fl} and \ref{BEL}, are plotted in figure 1. One
observes that the assumption that the GRs fully exhaust the EWSRs
is an overestimation. An appreciable part of the sum rule goes to
the excitation of the soft modes, specially for the case of E2
excitations. Indeed, for $^{8}\mathrm{B}$\ the EWSR given in eq.
\ref{EWSR} yield $S_{E1}^{total}=28$ $e^{2}$ fm$^{2}$ MeV and
$S_{E2}^{total}=890$ $e^{2}$ fm$^{4}$ MeV, respectively. These
values should be compared to the energy integrated multipole
response of the soft modes: $S_{E1}^{SD}=0.546$ $e^{2}$ fm$^{2}$
MeV and $S_{E2}^{SD}=396$ $e^{2}$ fm$^{4}$ MeV, respectively.
Although the E1 soft mode corresponds to a very small part of the
total sum-rule, it is responsible for large Coulomb dissociation
cross sections, since low energy E1 virtual photons are much more
abundant. In contrast, the E2 soft mode as obtained with the
potential model, exhausts 44\% of the EWSR. This is a hint that
the potential model overestimates the magnitude of the E2 response
function. Indeed, a recent experiment \cite{Dav98} has suggested
that the momentum distributions following the Coulomb breakup of
$^{8}\mathrm{B}$ can only be explained if the E2 response obtained
from the proton + $^{7}\mathrm{Be}$-core model is quenched by a
factor 2.
\begin{figure}
[t]
\begin{center}
\includegraphics[
height=2.4353in,
width=3.6019in
]%
{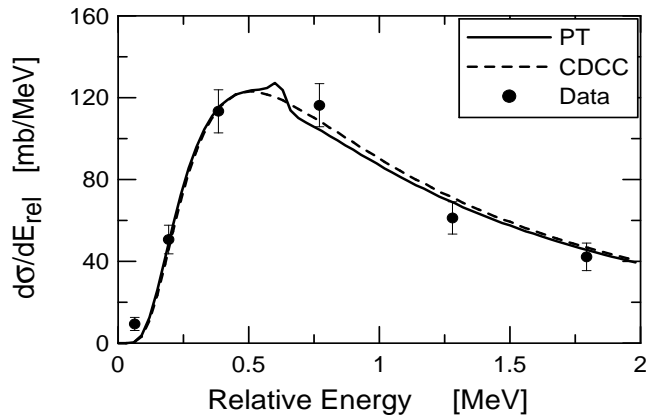}
\caption{Energy dependence of the cross section for $^{8}\mathrm{B}%
+\mathrm{Pb}\longrightarrow\mathrm{p}+^{7}\mathrm{Be}+\mathrm{Pb}$
at 84 MeV/nucleon. First-order perturbation calculations (PT) are
shown by the solid curve. The dashed curve is the result of a CDCC
calculation including the coupling between the ground state and
the low-lying states with the giant
dipole and quadrupole resonances. The data points are from ref. \cite{Da01}. }%
\end{center}
\end{figure}

The structure inputs as described above were used in a calculation using the
coupled-channels code DWEIKO \cite{BCG02} that includes relativistic dynamics,
important for bombarding energies of 84 MeV/nucleon. As the main interest is
for the low energy region, \ the energy mesh for the continuum discretization
was taken as 10 energy states equally spaced in the energy interval of 0 - 2
MeV. To account for the effect of the GR, 20 energy states equally spaced in
the interval 10 - 40 MeV were taken. The results of the coupled-channels
calculations were matched to results of first-order perturbation calculations
at impact parameters larger than 50 fm.

The results are plotted in figure 2. The solid curve shows the cross section
for the Coulomb dissociation reaction $^{8}\mathrm{B}+\mathrm{Pb}%
\longrightarrow\mathrm{p}+^{7}\mathrm{Be}+\mathrm{Pb}$ at 84 MeV/nucleon using
first-order perturbation (PT) theory \cite{CB96}. The numerical results have
been normalized to the data. The same normalization factor was used for the
coupled-channels results, shown by the dashed line. The first order
perturbation calculation shows a small peak at $E_{rel}=E_{x}-0.134$ MeV
$=640$ keV due to the excitation of the 1$^{+}$ resonance. This peak is
washed-out in the coupled-channels calculations due to the mesh size used. The
coupled-channels calculation is slightly different than the first-order
perturbation results, only for energies above 0.7 MeV. However, the correction
is very small, being no larger than 2\% for the whole energy interval.

We conclude that the effect of the giant resonances on the Coulomb
dissociation cross sections of $^{8}\mathrm{B}$ projectiles is
small and can be neglected. A similar conclusion is expected to
hold for the breakup reactions of other weakly-bound nuclei. The
total excitation cross sections of soft modes in $^{8}\mathrm{B}$\
for the reaction studied here are $\sigma_{E1}^{SD}=370$ mb
$\sigma_{E2}^{SD}=236$ mb, while the cross sections for the
excitation of GRs are  $\sigma_{GDR}=2.5$ mb \ and
$\sigma_{GQR}=6.5$ mb, respectively. This is the reason for the
small relevance of the GRs in the dynamic coupling. The situation
can be very different for the heavier nuclei which have a larger
response to the electromagnetic excitation in the region of giant
resonances. Although they have not yet been studied in details
experimentally, the electromagnetic response of heavy neutron- or
proton-rich nuclei close to the dripline will probably contain
soft multipole modes. These are very likely to be influenced by
the dynamic coupling to the much higher-lying giant resonance
states.

\section*{Acknowledgements}

I would like to thank Sam Austin for suggesting this problem and to Alex Brown
for providing the $^{8}\mathrm{B}$ density distribution parameters. This
research was supported in part by the U.S. National Science Foundation under
Grants No. PHY-007091 and PHY-00-70818.

\end{document}